\documentclass[10pt,twocolumn]{IEEEtran}
\usepackage[latin1]{inputenc}
\usepackage{amssymb}
\usepackage{latexsym}
\usepackage{mathrsfs}
\usepackage{amsfonts}
\usepackage{amsbsy}
\usepackage{amsmath}
\usepackage{times}
\usepackage{epsfig,epsf,times,amssymb,amsmath}
\usepackage{color}
\usepackage{setspace}
\usepackage{booktabs}
\usepackage{rotating}

\newcommand{\argmin}{\operatornamewithlimits{argmin}}
\newcommand{\argmax}{\operatornamewithlimits{argmax}}

\title{Adaptive Power Allocation Strategies using DSTC in Cooperative MIMO Networks}
\author{Tong Peng,~\IEEEmembership{Student Member,~IEEE,} Rodrigo C. de Lamare,~\IEEEmembership{Senior Member,~IEEE,}
and Anke Schmeink, ~\IEEEmembership{Member,~IEEE}
\thanks{Tong Peng, Rodrigo C. de Lamare are with Communications Research Group, Department of Electronics, University of York, York YO10 5DD, UK, e-mails: tp525@ohm.york.ac.uk; rcdl500@ohm.york.ac.uk}
\thanks{Anke Schmeink, Institute for Theoretical Information Technology, RWTH Aachen University, D-52056 Aachen, Germany, e-mail: schmeink@umic.rwth-aachen.de}}

\begin{document}
\maketitle
\IEEEpeerreviewmaketitle

\begin{abstract}
Adaptive Power Allocation (PA) algorithms with different criteria
for a cooperative Multiple-Input Multiple-Output (MIMO) network
equipped with Distributed Space-Time Coding (DSTC) are proposed and
evaluated. Joint constrained optimization algorithms to determine
the power allocation parameters, the channel parameters and the
receive filter are proposed for each transmitted stream in each
link. Linear receive filter and maximum-likelihood (ML) detection
are considered with Amplify-and-Forward (AF) and Decode-and-Forward
(DF) cooperation strategies. In the proposed algorithms, the
elements in the PA matrices are optimized at the destination node
and then transmitted back to the relay nodes via a feedback channel.
The effects of the feedback errors are considered. Linear MMSE
expressions and the PA matrices depend on each other and are updated
iteratively. Stochastic gradient (SG) algorithms are developed with
reduced computational complexity. Simulation results show that the
proposed algorithms obtain significant performance gains as compared
to existing power allocation schemes.
\end{abstract}

\section{Introduction}

Due to the benefits of cooperative multiple-input and
multiple-output (MIMO) systems \cite{Clarke}, extensive studies of
cooperative MIMO networks have been undertaken
\cite{P1}-\cite{Farhadi G.}. In \cite{P1}, an adaptive joint relay
selection and power allocation algorithm based on the minimum mean
square error (MMSE) criterion is designed. A joint transmit
diversity optimization and relay selection algorithm for the
Decode-and-Forward (DF) cooperating strategy \cite{J. N.
Laneman2004} is designed in \cite{P2}. A transmit diversity
selection matrix is introduced at each relay node in order to
achieve a better MSE performance by deactivating some relay nodes. A
central node which controls the transmission power for each link is
employed in \cite{O. Seong-Jun}. Although the centralized power
allocation can improve the performance significantly, the complexity
of the calculation increases with the size of the system. The works
on the power allocation problem for the DF strategy measuring the
outage probability in each relay node with a single antenna and
determining the power for each link between the relay nodes and the
destination node, have been reported in \cite{Min Chen}-\cite{Yindi
Jing}. The diversity gain can be improved by using relay nodes with
multiple antennas. When the number of relay nodes is the same, the
cooperative gain can be improved by using the DF strategy compared
with a system employing the AF strategy. However, the interference
at the destination will be increased if the relay nodes forward the
incorrectly detected symbols in the DF strategy. The power
allocation optimization algorithms in \cite{Ref1} and \cite{Ref2}
provide improved BER performance at the cost of requiring an
eigenvalue decomposition to obtain the key parameters.

In this paper, we propose joint adaptive power allocation (JAPA)
algorithms according to different optimization criteria with a
linear receiver or an ML detector for cooperative MIMO systems
employing multiple relay nodes with multiple antennas that perform
cooperating strategies. This work was first introduced and discussed
in \cite{Tong1} and \cite{Tong2}. The power allocation matrices
utilized in \cite{Tong1} are full rank and after the optimization,
all the parameters are transmitted back to the relay nodes and the
source node with an error-free and delay-free feedback channel. In
this paper, we employ the diagonal power allocation matrices in
which the parameters stand for the power allocated to each transmit
antenna. The requirement of the limited feedback is significantly
reduced as compared to the algorithms in the previously reported
works. It is worth to mention that the JAPA strategies derived in
our algorithms are two-phase optimization techniques, which
optimized the power assigned at the source node and at the relay
nodes in the first phase and the second phase iteratively, and the
proposed JAPA algorithms can be used as a power allocation strategy
for the second phase only.

Three optimization criteria, namely, MMSE, minimum bit error rate
(MBER) and maximum sum rate (MSR), are employed in the proposed JAPA
optimization algorithms in this paper. We firstly develop joint
optimization algorithms of the power allocation matrices and the
linear receive filter according to these three criteria,
respectively, which require matrix inversions and bring a high
computational burden to the receiver. In the proposed JAPA
algorithms with the MMSE, MBER, and MSR criteria, an SG method
\cite{S. Haykin} is employed in order to reduce the computational
complexity of the proposed algorithms. A comparison of the
computational complexity of the algorithms is considered in this
paper. A normalization procedure is employed by the optimization
algorithm in order to enforce the power constraint in both
transmission phases. After the normalization, the PA parameters are
transmitted back to each transmit node through a feedback channel.
The effect of the feedback errors is considered in the analysis and
in the simulation sections, where we indicate an increased MSE
performance due to feedback inaccuracy.

The paper is organized as follows. Section II introduces a two-hop
cooperative MIMO system with multiple relays applying the AF
strategy and the adaptive DSTC scheme. The constrained power
allocation problems for relay nodes and linear detection method are
derived in Section III, and the proposed JAPA SG algorithms are
derived in Section IV. Section V focuses on the computational
complexity comparison between the proposed and the existing
algorithms, and the effects of the feedback errors on the MSE of the
system. Section VI gives the simulation results and Section VII
provides the conclusion.

Notation: the italic, bold lower-case and bold upper-case letters
denote scalars, vectors and matrices, respectively. The operators
$E[\cdot]$ and $(\cdot)^\emph{H}$ stand for expected value and the
Hermitian operator. The $N \times N$ identity matrix is written as
${\boldsymbol I}_N$. $\parallel{\boldsymbol X}\parallel_F=\sqrt{{\rm
Tr}({\boldsymbol X}^H\cdot{\boldsymbol X})}=\sqrt{{\rm
Tr}({\boldsymbol X}\cdot{\boldsymbol X}^\emph{H})}$ is the Frobenius
norm. $\Re[\cdot]$ stands for the real part, and $Tr(\cdot)$ stands
for the trace of a matrix. $\rm{sgn}(\cdot)$ denotes the sign
function.

\section{Cooperative System Model}

\begin{figure}
\begin{center}
\def\epsfsize#1#2{0.825\columnwidth}
\epsfbox{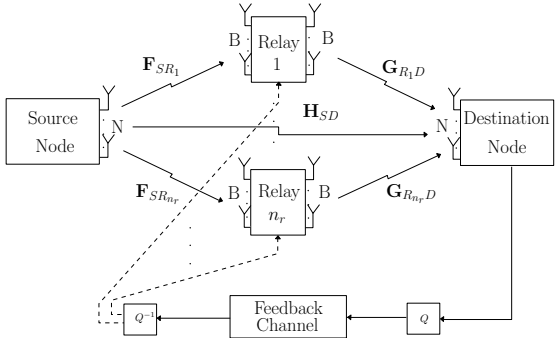}\vspace*{-1em} \caption{Cooperative MIMO
system model with $n_r$ relay nodes}\label{1}
\end{center}
\end{figure}

Consider a two-hop cooperative MIMO system in Fig. 1 with $n_r$
relay nodes that employs an AF cooperative strategy as well as a
DSTC scheme. The source node and the destination node have $N$
antennas to transmit and receive data. An arbitrary number of
antennas can be used at the relays which is denoted by $B$ shown in
Fig. 1. We consider only one user at the source node in our system
that operates in a spatial multiplexing configuration. Let
${\boldsymbol s}[i]$ denote the transmitted information symbol
vector at the source node which contains $N$ symbols ${\boldsymbol
s}[i] = [s_{1}[i], s_{2}[i], ... , s_{N}[i]]$, and has a covariance
matrix $E\big[{\boldsymbol s}[i]{\boldsymbol s}^{H}[i]\big] =
\sigma_{s}^{2}{\boldsymbol I}_N$, where $\sigma_s^2$ is the signal
power which we assume to be equal to 1. The source node broadcasts
${\boldsymbol s}[i]$ from the source to $n_r$ relay nodes as well as
to the destination node in the first hop, which can be described by
\begin{equation}\label{2.1}
\begin{aligned}
{\boldsymbol r}_{SD}[i] =& {\boldsymbol H}_{SD}[i]{\boldsymbol A}_S[i]{\boldsymbol s}[i] + {\boldsymbol n}_{SD}[i],\\
{\boldsymbol r}_{{SR}_{k}}[i]  =& {\boldsymbol F}_{SR_{k}}[i]{\boldsymbol A}_S[i]{\boldsymbol s}[i] + {\boldsymbol n}_{SR_k}[i],\\
 k =& 1,2,~...~ n_{r}, ~~i = 1,2,~...~,
\end{aligned}
\end{equation}
where ${\boldsymbol
A}_S[i]=\rm{diag}[a_{S_1}[i],a_{S_2}[i],...,a_{S_N}[i]]$ denotes the
diagonal $N \times N$ power allocation matrix assigned for the
source node, and ${\boldsymbol r}_{{SR}_{k}}[i]$ and ${\boldsymbol
r}_{SD}[i]$ denote the received symbol vectors at the $k$th relay
node and at the destination node, respectively. The $B \times 1$
vectors ${\boldsymbol n}_{{SR}_{k}}[i]$ and ${\boldsymbol
n}_{SD}[i]$ denote the zero mean complex circular symmetric additive
white Gaussian noise (AWGN) vector generated at the $k$th relay node
and at the destination node with variance $\sigma^{2}$. The matrices
${\boldsymbol F}_{SR_k}[i]$ and ${\boldsymbol H}_{SD}[i]$ are the $B
\times N$ channel coefficient matrices. It is worth to mention that
an orthogonal transmission protocol is considered which requires
that the source node does not transmit during the time period of the
second hop.

The received symbols are amplified and re-encoded at each relay node
prior to transmission to the destination node in the second hop. We
assume that the synchronization at each node is perfect. The
received vector ${\boldsymbol r}_{SR_k}[i]$ at the $k$th relay node
is assigned a $B \times B$ diagonal power allocation matrix
${\boldsymbol A}_k[i]=\rm{diag}[a_{k_1},a_{k_2},...,a_{k_B}]$ which
leads to $\tilde{{\boldsymbol s}}_{SR_k}[i]={\boldsymbol
A}_k[i]{\boldsymbol r}_{SR_k}[i]$. The $B \times 1$ signal vector
$\tilde{{\boldsymbol s}}_{SR_k}[i]$ will be re-encoded by a $B
\times T$ DSTC matrix ${\boldsymbol M}(\tilde{{\boldsymbol s}})$,
and then forwarded to the destination node. The relationship between
the $k$th relay and the destination node can be described as
\begin{equation}\label{2.2}
{\boldsymbol R}_{R_{k}D}[i] = {\boldsymbol G}_{R_kD}[i]{\boldsymbol M}_{R_{k}D}[i] + {\boldsymbol N}_{RD}[i].
\end{equation}
The $N \times T$ received symbol matrix ${\boldsymbol R}_{R_{k}D}[i]$ in
(\ref{2.2}) can be written as an $NT \times 1$ vector ${\boldsymbol
r}_{R_{k}D}[i]$ given by
\begin{equation}\label{2.3}
\begin{aligned}
{\boldsymbol r}_{R_{k}D}[i]  &= {\boldsymbol G}_{{eq}_k}[i]\tilde{{\boldsymbol s}}_{SR_k}[i] + {\boldsymbol n}_{R_{k}D}[i] = {\boldsymbol G}_{{eq}_k}[i]{\boldsymbol A}_k[i]{\boldsymbol r}_{SR_k}[i] + {\boldsymbol n}_{RD}[i]\\ &= {\boldsymbol G}_{{eq}_k}[i]{\boldsymbol A}_k[i]{\boldsymbol F}_{SR_{k}}[i]{\boldsymbol A}_S[i]{\boldsymbol s}[i] + {\boldsymbol n}_{R_k}[i] + {\boldsymbol n}_{RD}[i],
\end{aligned}
\end{equation}
where the $NT \times B$ matrix ${\boldsymbol G}_{eq_k}[i]$ stands
for the equivalent channel matrix which is the DSTC scheme
${\boldsymbol M}(\tilde{{\boldsymbol s}}[i])$ combined with the
channel matrix ${\boldsymbol G}_{R_{k}D}[i]$. The second term
${\boldsymbol n}_{R_k}[i]={\boldsymbol G}_{{eq}_k}[i]{\boldsymbol
A}_k[i]{\boldsymbol n}_{SR_k}[i]$ in (\ref{2.3}) stands for the
amplified noise received from the relay node, and the $NT \times 1$
equivalent noise vector ${\boldsymbol n}_{RD}[i]$ generated at the
destination node contains the noise parameters in ${\boldsymbol
N}_{RD}[i]$.

After rewriting ${\boldsymbol R}_{R_{k}D}[i]$ we can consider the
received symbol vector at the destination node as a $(T+1)N \times
1$ vector with two parts, one is from the source node and another
one is the superposition of the received vectors from each relay
node. Therefore, we can write the received symbol at the destination
node as
\begin{equation}\label{2.4}
\begin{aligned}
{\boldsymbol r}[i]  &= \left[\begin{array}{c} {\boldsymbol r}_{SD}[i] \\ {\boldsymbol r}_{RD}[i]\end{array} \right]=
\left[\begin{array}{c} {\boldsymbol H}_{SD}[i]{\boldsymbol A}_S[i]{\boldsymbol s}[i] + {\boldsymbol n}_{SD}[i] \\ \sum_{k=1}^{n_r}{\boldsymbol G}_{{eq}_k}[i]{\boldsymbol A}_k[i]{\boldsymbol F}_{SR_{k}}[i]{\boldsymbol A}_S[i]{\boldsymbol s}[i] + {\boldsymbol n}_{RD}[i] \end{array} \right] = \left[\begin{array}{c} {\boldsymbol H}_{eq_{SD}}[i] \\ \sum_{k=1}^{n_r}{\boldsymbol H}_{{eq}_k}[i] \end{array} \right]{\boldsymbol s}[i]+ \left[\begin{array}{c}{\boldsymbol n}_{SD}[i] \\ {\boldsymbol n}_{RD}[i] \end{array} \right] \\
& = {\boldsymbol H}_D[i]{\boldsymbol s}[i] + {\boldsymbol n}_D[i],
\end{aligned}
\end{equation}
where the $(T + 1)N \times N$ matrix ${\boldsymbol H}_D[i]$ denotes
the channel gain matrix with the power allocation of all the links
in the system. The $N \times N$ channel matrix ${\boldsymbol
H}_{eq_{SD}}[i]={\boldsymbol H}_{SD}[i]{\boldsymbol A}_S[i]$, while
the $k$th equivalent channel matrix ${\boldsymbol
H}_{{eq}_k}[i]={\boldsymbol G}_{{eq}_k}[i]{\boldsymbol
A}_k[i]{\boldsymbol F}_{SR_{k}}[i]{\boldsymbol A}_S[i]$. We assume
that the coefficients in all channel matrices are statistically
independent and remain constant over the transmission. The $(T + 1)N
\times 1$ noise vector ${\boldsymbol n}_D[i]$ contains the
equivalent received noise vector at the destination node, which can
be modeled as AWGN with zero mean and covariance matrix
$\sigma^{2}(1+\parallel\sum_{k=1}^{n_r}{\boldsymbol
G}_{{eq}_k}[i]{\boldsymbol A}_k[i]\parallel^2_F){\boldsymbol
I}_{(T+1)N}$. It is worth to mention that the value in $B$ is
variable and, in this work, we focus on the power allocation
optimization algorithms in cooperative MIMO systems. For simplicity,
we consider scenarios in which there are $B=N$ antennas at the
relays.

\section{Adaptive Power Allocation Matrix Optimization Strategies}

In this section, we consider the design of a two-phase adjustable
power allocation matrix according to various criteria using a DSTC
scheme in cooperative MIMO systems. The linear receive filter is
determined jointly with the power allocation matrices. A feedback
channel is considered in order to convey the information about the
power allocation prior to transmission to the destination node.

\subsection{Joint Linear MMSE Receiver Design with Power Allocation}

The linear MMSE receiver design with power allocation matrices is
derived as follows. By defining the $(T+1)N \times 1$ parameter
vector ${\boldsymbol w}_j[i]$ to determine the $j$th symbol $s_j[i]$
in the signal vector ${\boldsymbol s}[i]$, we propose the MSE based
optimization with a power constraint described by
\begin{equation}\label{3.1}
\begin{aligned}
    &\left[{\boldsymbol w}_j[i],{\boldsymbol A}_S[i],{\boldsymbol A}_k[i]]\right] = \argmin_{{\boldsymbol w}_j[i],{\boldsymbol A}_S[i],{\boldsymbol A}_k[i]} E\left[\|s_j[i]-{\boldsymbol w}_j^\emph{H}[i]{\boldsymbol r}[i]\|^2\right],\\ & ~~~~~~~~~s.t.~
    \rm{Tr}(\sum_{k=1}^{n_r}{\boldsymbol A}_k[i]{\boldsymbol A}_k^\emph{H}[i])\leq \rm{P_R},~\rm{Tr}({\boldsymbol A}_S[i]{\boldsymbol A}_S^\emph{H}[i])\leq \rm{P_T},
\end{aligned}
\end{equation}
where $P_T$ and $P_R$ denote the transmit power assigned to all the
relay nodes and to the source node, respectively. The values of the
parameters in the power allocation (PA) matrices are restricted by
$P_T$ and $P_R$. By employing the Lagrange multipliers $\lambda_1$
and $\lambda_2$ we can obtain the Lagrangian function shown as
\begin{equation}\label{3.2}
    \mathscr{L} = E\left[\|s_j[i]-{\boldsymbol w}_j^\emph{H}[i]{\boldsymbol r}[i]\|^2\right]+\lambda_1(\sum_{j=1}^{N}a_{S_j}[i]-P_T)+\lambda_2(\sum_{k=1}^{n_r}\sum_{j=1}^{N}a_{k_j}[i]-P_R),
\end{equation}
where $a_{S_j}[i]$ denotes the $j$th parameters in the diagonal of
${\boldsymbol A}_S[i]$ while $a_{k_j}[i]$ stands for the $j$th
parameters in the diagonal of ${\boldsymbol A}_k[i]$.

By expanding the right-hand side of (\ref{3.2}), taking the gradient
with respect to ${\boldsymbol w}_j^*[i]$, $a^*_{S_j}[i]$ and
$a^*_{k_j}[i]$, respectively, and equating the terms to zero, we can
obtain
\begin{equation}\label{3.3}
\begin{aligned}
    {\boldsymbol w}_j[i]&={\boldsymbol R}^{-1}{\boldsymbol p},\\
    a_{S_j}[i] & = \tilde R^{-1}_S\tilde P_S,\\
     a_{k_j}[i] & = \tilde R^{-1}\tilde P,
\end{aligned}
\end{equation}
where
\begin{equation}\label{3.4.1}
\begin{aligned}
    {\boldsymbol R} &=E\left[{\boldsymbol r}[i]{\boldsymbol r}^\emph{H}[i]\right],~{\boldsymbol p}=E\left[{\boldsymbol r}[i]s_j^*[i]\right],\\
    \tilde R_S & = E\left[{\boldsymbol w}^\emph{H}_j[i]{\boldsymbol h}_{{SDA}_j}[i]s_j[i]s^*_j[i]{\boldsymbol h}^\emph{H}_{{SDA}_j}[i]{\boldsymbol w}_j[i] + \lambda_1 a_{S_j}[i]\right],\\
    \tilde P_S & = E\left[{\boldsymbol h}^\emph{H}_{{SDA}_j}[i]{\boldsymbol w}_j[i]s^*_j[i]s_j[i]\right],\\
    \tilde P & = E\left[s_j[i]s^*_j[i]a_{S_j}^*[i]f^*_{k_j}[i]{\boldsymbol g}^\emph{H}_{eq_{k_j}}[i]{\boldsymbol w}_j[i]\right],\\
    \tilde R & = E\left[{\boldsymbol w}^\emph{H}_j[i]{\boldsymbol g}_{eq_{k_j}}[i]f_{k_j}[i]a_{S_j}[i]s_j[i]s^*_j[i]a_{S_j}^*[i]f^*_{k_j}[i]{\boldsymbol g}^\emph{H}_{eq_{k_j}}[i]{\boldsymbol w}_j[i]+\lambda_2 a_{k_j}[i]\right].
\end{aligned}
\end{equation}
The vector ${\boldsymbol h}_{SDA_j}$ denotes the channel vector
assigned to the parameter $a_{S_j}$ and is the $j$th column of the
equivalent channel matrix ${\boldsymbol H}_{SDA}[i] = \left[
{\boldsymbol H}_{SD}[i] ; {\boldsymbol G}_{{eq}_k}[i]{\boldsymbol
A}_k[i]{\boldsymbol F}_{SR_k}[i] \right]$ , and $f_{k_j}[i]$ and
${\boldsymbol g}_{eq_{k_j}}[i]$ denotes the $j$th parameter in
${\boldsymbol F}_k[i]$ and the $j$th column in ${\boldsymbol
G}_{eq_k}[i]$, respectively. The value of the Lagrange multipliers
$\lambda_1$ and $\lambda_2$ can be determined by substituting
${\boldsymbol A}_S[i]$ and ${\boldsymbol A}_k[i]$ into
$\rm{Tr}({\boldsymbol A}_S[i]{\boldsymbol A}_S^\emph{H}[i])\leq
\rm{P_T}$ and $\lambda\rm{Tr}(\sum_{k=1}^{n_r}{\boldsymbol
A}_k[i]{\boldsymbol A}^\emph{H}_k[i])= \rm{P_R}$, respectively, and
then solving the power constraint equations. The problem is that a
high computational complexity of ${\rm O}(((T+1)N)^3)$ is required,
and it will increase cubically with the number of antennas or the
use of more complicated STC encoders.

\subsection{Joint Linear MBER Receiver Design with Power Allocation}

The MBER receiver design \cite{delamare_mber,Shengchen} with power
allocation in the second phase is derived as follows. The BPSK
modulation scheme is utilized for simplicity. According to the
expression in (\ref{2.4}), the desired information symbols at the
destination node can be computed as
\begin{equation}\label{3.5}
b_j[i] = \rm{sgn}({\boldsymbol w}^\emph{H}_j[i]{\boldsymbol r}[i])=\rm{sgn}(\tilde{s}_j[i]),
\end{equation}
where $\tilde{s}_j[i]$ denotes the detected symbol at the receiver
which can be further written as
\begin{equation}\label{3.6}
\begin{aligned}
    \tilde{s}_j[i] & = \Re\left[{\boldsymbol w}^\emph{H}_j[i]{\boldsymbol r}[i]\right]
     = \Re\left[{\boldsymbol w}^\emph{H}_j[i]({\boldsymbol H}_D[i]{\boldsymbol s}[i] + {\boldsymbol n}_D[i])\right]
    = \Re\left[{\boldsymbol w}^\emph{H}_j[i]{\boldsymbol H}_D[i]{\boldsymbol s}[i] + {\boldsymbol w}^\emph{H}_j[i]{\boldsymbol n}_D[i]\right]\\
    & = \Re\left[\tilde{s}'_j[i] + e_j[i]\right],
\end{aligned}
\end{equation}
where $\tilde{s}'_j[i]$ is the noise-free detected symbol, and
$e_j[i]$ denotes the error factor for the $j$th detected symbol.
Define an $N \times N_b$ matrix $\boldsymbol {\bar{S}}$ which is
constructed by a set of vectors ${\boldsymbol
{\bar{s}}}_l=[s_{l_1},s_{l_2},...,s_{l_N}]^T,~l = 1,2,...,N_b$ and
$N_b=2^N$, containing all the possible combinations of the
transmitted symbol vector ${\boldsymbol s}[i]$ and we can obtain
\begin{equation}\label{3.7}
    {\bar s}_{l_j}[i] = \Re\left[{\boldsymbol w}^\emph{H}_j[i]{\boldsymbol H}_D[i]{\boldsymbol {\bar{s}}}_l + {\boldsymbol w}^\emph{H}_j[i]{\boldsymbol n}_D[i]\right] = \Re\left[{\boldsymbol w}^\emph{H}_j[i]\boldsymbol {\bar{r}}_l[i]]+ e_{l_j}[i]\right]=\bar{s}'_{l_j}[i] + e_{l_j}[i],
\end{equation}
where $\bar{s}'_{l_j}[i]$ denotes the noise-free detected symbol in
the $l$th column and the $j$th row of ${\boldsymbol {\bar{S}}}$.
Since the probability density function (pdf) of $\boldsymbol
{\bar{r}}[i]$ is given by
\begin{equation}\label{3.8}
    p_{\boldsymbol {\bar{r}}[i]} = \frac{1}{N_b\sqrt{2\pi\sigma^2_{n}{\boldsymbol w}^\emph{H}_j[i]{\boldsymbol w}_j[i]}}\sum_{l=1}^{N_b}\exp\big(-\frac{(\bar{s}_{l_j}[i]-\bar{s}'_{l_j}[i])^2}{2\sigma^2_{n}{\boldsymbol w}^\emph{H}_j[i]{\boldsymbol w}_j[i]}\big),
\end{equation}
by employing the $Q$ function, we can obtain the BER expression of
the cooperative MIMO system which is
\begin{equation}\label{3.9}
    P_E({\boldsymbol w}_j[i],a_{S_j}[i],a_{k_j}[i])=\frac{1}{N_b}\sum^{N_b}_{l=1}{\boldsymbol Q}(c_{l_j}[i]),
\end{equation}
where
\begin{equation}\label{3.10}
    c_{l_j}[i] = \frac{sgn(s_{l_j})\bar{s}_{l_j}'[i]}{\sigma_{n}\sqrt{{\boldsymbol w}^\emph{H}_j[i]{\boldsymbol w}_j[i]}}=\frac{sgn(s_{l_j})\Re\left[{\boldsymbol w}^\emph{H}_j[i]\boldsymbol {\bar{r}}_l[i]\right]}{\sigma_{n}\sqrt{{\boldsymbol w}^\emph{H}_j[i]{\boldsymbol w}_j[i]}}.
\end{equation}
The joint power allocation with linear receiver design problem is
given by
\begin{equation}\label{3.11}
\begin{aligned}
    &[{\boldsymbol w}_j[i],a_{S_j}[i],a_{k_j}[i]] = \argmin_{{\boldsymbol w}_j[i],a_{S_j}[i],a_{k_j}[i]}P_E({\boldsymbol w}_j[i],a_{S_j}[i],a_{k_j}[i]) ,\\ & ~~~~~~~~~s.t.~
    \sum_{j=1}^{N}a_{S_j}[i]\leq \rm{P_T},~\sum_{k=1}^{n_r}\sum_{j=1}^{N}a_{k_j}[i]\leq \rm{P_R}.
\end{aligned}
\end{equation}
According to (\ref{3.9}) and (\ref{3.10}), the solution of the
design problem in (\ref{3.11}) with respect to ${\boldsymbol
w}_j[i]$, $a_{S_j}[i]$ and $a_{k_j}[i]$ is not a closed-form one.
Therefore, we design an adaptive JAPA strategy according to the MBER
criterion using the SG algorithm in order to update the parameters
iteratively to achieve the optimal solution in the next section.

\subsection{Joint Linear MSR Receiver Design with Power Allocation}

We will develop a joint power allocation strategy focuses on
maximizing the sum rate at the destination node. The expression of
the sum rate after the detection is derived in \cite{TongW} as
\begin{equation}\label{3.12}
    I = \frac{1}{2}\log_2(1+SNR_{ins}),
\end{equation}
where
\begin{equation}\label{3.13}
    SNR_{ins} = \frac{E[{\boldsymbol s}^\emph{H}{\boldsymbol s}]Tr({\boldsymbol W}^\emph{H}[i]{\boldsymbol H}_D[i]{\boldsymbol H}^\emph{H}_D[i]{\boldsymbol W}[i])}{E[{\boldsymbol n}^\emph{H}_D[i]{\boldsymbol n}_D[i]]},
\end{equation}
and ${\boldsymbol W}[i]=[{\boldsymbol w}_1[i],{\boldsymbol w}_2[i],...,{\boldsymbol w}_N[i]]$ denotes the $N(T+1) \times N$ linear receive filter matrix, and ${\boldsymbol n}[i]$ denotes the received noise vector. By substituting (\ref{2.4}) into (\ref{3.13}), we can obtain
\begin{equation}\label{3.14}
    SNR_{ins} = \frac{\sigma^2_sTr({\boldsymbol W}^\emph{H}[i](\sum_{k=1}^{n_r}{\boldsymbol G}_{{eq}_k}[i]{\boldsymbol A}_k[i]{\boldsymbol F}_{SR_{k}}[i]{\boldsymbol A}_S[i])(\sum_{k=1}^{n_r}{\boldsymbol G}_{{eq}_k}[i]{\boldsymbol A}_k[i]{\boldsymbol F}_{SR_{k}}[i]{\boldsymbol A}_S[i])^\emph{H}{\boldsymbol W}[i])}{\sigma^2_nTr({\boldsymbol W}^\emph{H}[i]({\boldsymbol I}_{N(T+1)}+(\sum_{k=1}^{n_r}{\boldsymbol G}_{{eq}_k}[i]{\boldsymbol A}_{R_kD}[i])(\sum_{k=1}^{n_r}{\boldsymbol G}_{{eq}_k}[i]{\boldsymbol A}_{R_kD}[i])^\emph{H}){\boldsymbol W}[i])}.
\end{equation}
Since the logarithm is an increasing function, maximizing the sum
rate is equivalent to maximizing the instantaneous SNR. The
optimization problem can be written as
\begin{equation}\label{3.16}
\begin{aligned}
    \left[{\boldsymbol W}[i],{\boldsymbol A}_S[i],{\boldsymbol A}_k[i]\right]=\argmax_{{\boldsymbol W}[i],{\boldsymbol A}_S[i],{\boldsymbol A}_k[i]}SNR_{ins},
    ~s.t.~\rm{Tr}({\boldsymbol A}_k[i]{\boldsymbol A}_k^\emph{H}[i])\leq \rm{P_R},~\rm{Tr}({\boldsymbol A}_S[i]{\boldsymbol A}_S^\emph{H}[i])\leq \rm{P_T},
\end{aligned}
\end{equation}
where $SNR_{ins}$ is given by (\ref{3.14}).

As expressed in (\ref{3.14}), the solution of (\ref{3.16}) with
respect to the matrices ${\boldsymbol W}[i]$, ${\boldsymbol A}_S[i]$
and ${\boldsymbol A}_k[i]$ does not result in closed-form
expressions. Therefore, in the next section we propose a JAPA SG
algorithm to obtain the joint optimization algorithm for determining
the linear receiver filter parameters and power allocation matrices
to maximize the sum rate.

\section{Low Complexity Joint Linear Receiver Design with Power Allocation}

In this section, we jointly design an adjustable power allocation
matrix and the linear receiver for the DSTC scheme in cooperative
MIMO systems. Adaptive SG algorithms \cite{S. Haykin} with reduced
complexity are devised.

\subsection{Joint Adaptive SG Estimation for MMSE Receive Filter and Power Allocation}

According to (\ref{3.1}) and (\ref{3.2}), the joint optimization
problem for power allocation matrices and receiver parameter vectors
depend on each other. By computing the instantaneous gradient terms
of (\ref{3.2}) with respect to ${\boldsymbol w}_j[i]$, $a_{S_j}[i]$
and $a_{k_j}[i]$, respectively, we can obtain
\begin{equation}\label{4.1}
\begin{aligned}
    \nabla\mathscr{L}_{{\boldsymbol w}^*_j[i]} & =  -{\boldsymbol r}[i](s_j[i] - {\boldsymbol w}^\emph{H}_j[i]{\boldsymbol r}[i]))^*  =  -{\boldsymbol r}[i]e^*_j[i],\\
    \nabla\mathscr{L}_{a^*_{S_j}[i]} & = -\nabla_{a^*_{S_j}[i]}({\boldsymbol w}^\emph{H}_j[i]{\boldsymbol h}_{{SDA}_j}[i]a_{S_j}[i]s_j[i])^\emph{H}(s_j[i]- {\boldsymbol w}^\emph{H}_j[i]{\boldsymbol r}[i])
    = {\boldsymbol h}^\emph{H}_{{SDA}_j}[i]{\boldsymbol w}_j[i]s^*_j[i]e_j[i],\\
    \nabla\mathscr{L}_{a^*_{k_j}[i]} & = -\nabla_{a^*_{k_j}[i]}({\boldsymbol w}^\emph{H}_{R_j}[i]{\boldsymbol g}_{eq_{k_j}}[i]{\boldsymbol f}_{k_j}[i]a_{k_j}[i]{\boldsymbol s}[i])^\emph{H}(s_j[i]- {\boldsymbol w}^\emph{H}_j[i]{\boldsymbol r}[i]) =  -({\boldsymbol g}_{eq_{k_j}}[i]{\boldsymbol f}_{k_j}[i]{\boldsymbol s}[i])^\emph{H}{\boldsymbol w}_{R_j}[i]e_j[i],
\end{aligned}
\end{equation}
where ${\boldsymbol h}_{{SDA}_j}[i]$ denotes the $j$th column with
dimension $N(T+1) \times 1$ of the equivalent channel matrix
${\boldsymbol H}_{SDA}[i]$, and ${\boldsymbol g}_{{eq_k}_j}[i]$ and
${\boldsymbol f}_{k_j}[i]$ denote the $j$th column and the $j$th row
of the channel matrices ${\boldsymbol F}_k[i]$ and ${\boldsymbol
G}_{eq_{k_j}}[i]$, respectively. The $NT \times 1$ vector
${\boldsymbol w}_{R_j}[i]$ is the parameter vector for the received
symbols from the relay nodes. The error signal is denoted by
$e_j[i]=s_j[i] - {\boldsymbol w}_j^H[i]{\boldsymbol r}[i]$. We can
devise an adaptive SG estimation algorithm by using the
instantaneous gradient terms of the Lagrangian which were previously
derived with SG descent rules \cite{S. Haykin}:
\begin{equation}\label{4.2}
\begin{aligned}
    {\boldsymbol w}_j[i + 1] & = {\boldsymbol w}_j[i] - \mu\nabla\mathscr{L}_{{\boldsymbol w}^*_j[i]},\\
    a_{S_j}[i + 1] & = a_{S_j}[i] - \nu\nabla\mathscr{L}_{a^*_{S_j}[i]},\\
    a_{k_j}[i + 1] & = a_{k_j}[i] - \tau\nabla\mathscr{L}_{a^*_{k_j}[i]},
\end{aligned}
\end{equation}
where $\mu$, $\nu$ and $\tau$ are the step sizes of the recursions
for the estimation procedure. The computational complexity of
${\boldsymbol w}_j[i]$, $a_{S_j}[i]$ and $a_{k_j}[i]$ in (\ref{4.2})
is $(\boldsymbol O(NT))$, $(\boldsymbol O(3NT))$ and $(\boldsymbol
O(N^2T^2))$, respectively, which is much less than that of the
algorithm we described in Section III.

It is worth to mention that instead of calculating the Lagrange
multiplier $\lambda$, a normalization of the power allocation
matrices after the optimization which ensures that the energy is not
increased is required and implemented as
\begin{equation}\label{4.2.1}
\begin{aligned}
  {\boldsymbol A}_S[i+1] =& \frac{\sqrt{\rm{P_T}}{\boldsymbol A}_S[i+1]}{\parallel{\boldsymbol A}_S[i+1]\parallel_F},\\
  {\boldsymbol A}_k[i+1] =& \frac{\sqrt{\rm{P_R}}{\boldsymbol A}_k[i+1]}{\parallel\sum_{k=1}^{n_r}\rm{Tr}({\boldsymbol A}_k[i+1]\parallel_F}.
\end{aligned}
\end{equation}

\subsection{Joint Adaptive MBER SG Estimation and Power Allocation}

The key strategy to derive an adaptive estimation algorithm for
solving (\ref{3.11}) is to find out an efficient and reliable method
to calculate the pdf of the received symbol vector ${\boldsymbol
r}[i]$ at the destination node. According to the algorithms in
\cite{Bowman}, kernel density estimation provides an effective
method for accurately estimating the required pdf.

By transmitting a block of $M$ training samples ${\boldsymbol {\hat{s}}}=\rm{sgn}({\boldsymbol {\hat{b}}})$, the kernel density estimated pdf of ${\boldsymbol {\hat{s}}}[i]$ is given by
\begin{equation}\label{4.3}
    p_{\hat{s}} = \frac{1}{M\sqrt{2\pi}\rho_n\sqrt{{\boldsymbol w}^\emph{H}_j[i]{\boldsymbol w}_j[i]}}\sum_{j=1}^{M}exp\big(-\frac{(\tilde{s}_j-\hat{s}_j)^2}{2\rho^2_n{\boldsymbol w}^\emph{H}_j[i]{\boldsymbol w}_j[i]}\big),
\end{equation}
where $\rho_n$ is related to the standard deviation of noise $\sigma_n$ and it is suggested in \cite{Bowman} that a lower bound of $\rho_n=\big(\frac{4}{3M}\big)^{\frac{1}{5}}\sigma_n$ should be chosen. The symbol $\tilde{s}_j$ is calculated by (\ref{3.7}), and $\hat{s}_j$ stands for the $j$th element in the $M \times 1$ training samples ${\boldsymbol {\hat{s}}}$. The expression of the BER can be derived as
\begin{equation}\label{4.4}
    \hat{P_E}({\boldsymbol w}_j[i],a_{S_j}[i],a_{k_j}[i])=\frac{1}{M}\sum^{M}_{j=1}{\boldsymbol Q}(c_j[i]),
\end{equation}
where
\begin{equation}\label{4.5}
    c_j[i] = \frac{sgn(\hat{s}_j)\bar{\hat{s}}_j}{\rho_n\sqrt{{\boldsymbol w}^\emph{H}_j[i]{\boldsymbol w}_j[i]}}.
\end{equation}

By substituting (\ref{4.5}) into (\ref{4.4}) and taking the gradient with respect to different arguments, we can obtain
\begin{equation}\label{4.6}
    \nabla P_{E_{{\boldsymbol w}_j}}[i]  = \frac{1}{M\sqrt {2\pi}\sqrt{{\boldsymbol w}^\emph{H}_j[i]{\boldsymbol w}_j[i]}}\sum^{M}_{j=1}\exp\big(-\frac{c^2_j[i]}{2}\big)sgn(s_j)
    \frac{{\boldsymbol{\bar{r}}}[i] - \frac{1}{2}\bar{s}_j[i]{\boldsymbol w}_j[i]}{\sigma_{n}{\boldsymbol w}^\emph{H}_j[i]{\boldsymbol w}_j[i]},
\end{equation}
\begin{equation}\label{4.7}
    \nabla P_{E_{a_{S_j}}}[i] = \frac{1}{M\sqrt{2\pi}\sigma_{n}\sqrt{{\boldsymbol w}^\emph{H}_j[i]{\boldsymbol w}_j[i]}}\sum^{M}_{j=1}exp\big(-\frac{c^2_j[i]}{2}\big)sgn(s_j)\Re[{\boldsymbol w}^\emph{H}_j[i]{\boldsymbol h}_{SDA_j}[i]s_j],
\end{equation}
\begin{equation}\label{4.8}
    \nabla P_{E_{a_{k_j}}}[i] = \frac{1}{M\sqrt{2\pi}\sigma_{n}\sqrt{{\boldsymbol w}^\emph{H}_{D_j}[i]{\boldsymbol w}_{R_j}[i]}}\sum^{M}_{j=1}exp\big(-\frac{c^2_j[i]}{2}\big)sgn(s_j)\Re[{\boldsymbol w}^\emph{H}_{R_j}[i]{\boldsymbol h}_{k_j}[i]s_j],
\end{equation}
where ${\boldsymbol h}_{k_j}[i]={\boldsymbol
g}_{eq_{k_j}}[i]f_{k_j}[i]$ denotes the equivalent channel vector
assigned for $s_{m_j}$. By making use of an SG algorithm in \cite{S.
Haykin}, the updated ${\boldsymbol w}_j[i]$, $a_{S_j}[i]$
and$a_{k_j}[i]$ can be calculated by (\ref{4.2}). The convergence
property of the joint iterative optimization problems have been
tested and proved by Niesen et al. in \cite{UNiesen}. In the
proposed design problem the receive filter parameter vectors and the
power allocation parameters depend on each other, and the proposed
JAPA algorithms provide an iterative update process and finally both
of the desired items will reach at least a local optimum of the BER
cost function.

\subsection{Joint Adaptive MSR SG Algorithm for Power Allocation and Receiver Design}

The proposed power allocation algorithm that maximizes the sum rate
at the destination node is derived as follows. We consider the
design problem in (\ref{3.16}) and the instantaneous received
$SNR_{ins}$ as given in (\ref{3.14}). According to the property of
the trace $\rm{Tr}(\cdot)$ we can obtain
\begin{equation}\label{4.9}
    SNR_{ins} = \frac{\sigma^2_s}{n_{eq}[i]}Tr({\boldsymbol R}_{SDA}[i]{\boldsymbol A}^\emph{H}_S[i])
    = \frac{\sigma^2_s}{n_{eq}[i]}Tr(\sum_{k=1}^{n_r}{\boldsymbol R}_{{\boldsymbol G}_{eq_k}}[i]{\boldsymbol A}^\emph{H}_k[i]),
\end{equation}
where
\begin{equation*}
\begin{aligned}
    & {\boldsymbol R}_{SDA}[i] = {\boldsymbol H}^\emph{H}_{SDA}[i]{\boldsymbol W}[i]{\boldsymbol W}^\emph{H}[i]{\boldsymbol H}_{SDA}[i]{\boldsymbol A}_S[i],\\
    & {\boldsymbol R}_{{\boldsymbol G}_{eq_k}}[i] = {\boldsymbol G}^\emph{H}_{eq_{k}}[i]{\boldsymbol W}[i]{\boldsymbol W}^\emph{H}[i]{\boldsymbol H}_{SDA}[i]{\boldsymbol A}_S[i]{\boldsymbol A}_S^\emph{H}[i]{\boldsymbol F}_{SR_k}^\emph{H}[i],\\
    & n_{eq}[i] = \sigma^2_nTr({\boldsymbol W}^\emph{H}[i]{\boldsymbol W}[i]+{\boldsymbol W}^\emph{H}[i](\sum_{k=1}^{n_r}{\boldsymbol G}_{eq_k}[i]{\boldsymbol A}_k[i])(\sum_{k=1}^{n_r}{\boldsymbol G}_{eq_k}[i]{\boldsymbol A}_k[i])^\emph{H}{\boldsymbol W}[i]).
\end{aligned}
\end{equation*}

Since the power allocation matrices ${\boldsymbol A}_S[i]$ and ${\boldsymbol A}_k[i]$ are diagonal, we just focus on the terms containing the conjugate of the $j$th parameter in order to simplify the derivation, and rewrite (\ref{4.9}) as
\begin{equation}\label{4.10}
    SNR_{ins} = \frac{\sigma^2_s\sum_{k=1}^{n_r}\sum_{j=1}^{N}r_{SDA_j}[i]a^*_{S_j}[i]}{\sigma^2_n\sum_{j=1}^{N}({\boldsymbol w}^\emph{H}_j[i]{\boldsymbol w}_j[i]+\sum_{k=1}^{n_r}r_{{{\boldsymbol N}_k}_j}[i]a^*_{k_j}[i])}=\frac{\sigma^2_s\sum_{k=1}^{n_r}\sum_{j=1}^{N}r_{{\boldsymbol G}_{eq_{k_j}}}[i]a^*_{k_j}[i]}{\sigma^2_n\sum_{j=1}^{N}({\boldsymbol w}^\emph{H}_j[i]{\boldsymbol w}_j[i]+\sum_{k=1}^{n_r}r_{{{\boldsymbol N}_k}_j}[i]a^*_{k_j}[i])},
\end{equation}
where $r_{{\boldsymbol G}_{eq_{k_j}}}[i]$ and $r_{{{\boldsymbol N}_k}_j}[i]$ denotes the $j$th element in the diagonal of ${\boldsymbol R}_{{\boldsymbol G}_{eq_k}}[i]$ and ${\boldsymbol R}_{{\boldsymbol N}_k}[i]$, respectively. ${\boldsymbol R}_{{\boldsymbol N}_k}[i]={\boldsymbol G}^\emph{H}_{eq_{k}}[i]{\boldsymbol W}[i]{\boldsymbol W}^\emph{H}[i]{\boldsymbol G}_{eq_k}[i]{\boldsymbol A}_k[i]$ denotes the equivalent matrix assigned for the noise at the $k$th relay node.

By taking the stochastic gradient of (\ref{4.10}) with respect to $a^*_{S_j}[i]$, $a^*_{k_j}[i]$ and ${\boldsymbol W }^\emph{H}[i]$ we can obtain
\begin{equation}\label{4.11}
\begin{aligned}
    \nabla_{{\boldsymbol W}[i]} & = \frac{\sigma^2_s}{n_{eq}[i]}(Tr(\parallel{\boldsymbol H}_{SDA}[i]{\boldsymbol A}_S[i]\parallel^2_F{\boldsymbol W}[i])n_{eq}[i]\\ &~~~~~~~-\parallel{\boldsymbol W}^\emph{H}[i]{\boldsymbol H}_{SDA}[i]{\boldsymbol A}_S[i]\parallel^2_F Tr({\boldsymbol W}[i]+{\boldsymbol G}_{eq_k}[i]{\boldsymbol A}_k[i]({\boldsymbol G}_{eq_k}[i]{\boldsymbol A}_k[i]){\boldsymbol W}[i])),\\
    \nabla_{a_{S_j}[i]} & = \frac{\sigma^2_s}{n_{eq}[i]}r_{SDA_j}[i],\\
    \nabla_{a_{R_kD}[i]} & = \frac{\sigma^2_s}{n_{eq}[i]}(r_{{\boldsymbol G}_{eq}}[i]n_{eq}[i] - \sigma^2_nr_{{{\boldsymbol N}_k}_j}[i]\sum_{j=1}^{N}r_{{\boldsymbol G}_{eq_{k_j}}}[i]a^*_{k_j}[i]).
\end{aligned}
\end{equation}
By using (\ref{4.2}) and (\ref{4.2.1}) the proposed algorithm is achieved. Table I shows a summary of the JAPA SG algorithms with different criteria. A low complexity channel estimation method derived in \cite{Tong1} can be also employed to obtain the channel matrices required in the proposed algorithms.

\begin{table}[!t]
\centering
    \caption{The JAPA SG Algorithms}     
    \label{tab:JAPA-SG}
    \begin{small}
        \begin{tabular}{|l|}
\hline

1: \bfseries{Initialization}: \\
 \qquad ${\boldsymbol W}[0]={\boldsymbol I}_{(T+1)N \times 1}$,\\
 \qquad $a_{S_j}[0] = 1, ~a_{k_j}[0]=1$, \\
 \qquad ${\boldsymbol H}_{SDA}[i]=\sum_{k=1}^{n_r}{\boldsymbol G}_{eq_k}[i]{\boldsymbol A}_k[i]{\boldsymbol F}_{SR_k}[i]$,\\
 \hline

2:  \bfseries {for} $j=1$ \bfseries{to} $N$ \bfseries {do} \\

\qquad 2-1: \bfseries{JAPA SG MMSE Algorithm} \\
 \qquad \qquad $e_j[i]=s_j[i] - {\boldsymbol w}^\emph{H}_j[i]{\boldsymbol r}[i],$ \\
 \qquad \qquad $\nabla\mathscr{L}_{{\boldsymbol w}^\emph{H}_j[i]} = -{\boldsymbol r}[i]e^*_j[i],$\\
 \qquad \qquad $\nabla\mathscr{L}_{a^*_{S_j}[i]} = {\boldsymbol h}^\emph{H}_{{SDA}_j}[i]{\boldsymbol w}_j[i]s^*_j[i]e_j[i],$\\
 \qquad \qquad $\nabla\mathscr{L}_{a^*_{k_j}[i]} = -({\boldsymbol g}_{eq_{k_j}}[i]{\boldsymbol f}_{k_j}[i]{\boldsymbol s}[i])^\emph{H}{\boldsymbol w}_{D_j}[i]e_j[i],$ \\
 \hline

 \qquad 2-2: \bfseries{JAPA SG MBER Algorithm} \\
 \qquad \qquad $c_{m_j}[i] = \frac{sgn(\hat{s}_{m_j})\bar{\hat{s}}_{m_j}}{\rho_n\sqrt{{\boldsymbol w}^\emph{H}_j[i]{\boldsymbol w}_j[i]}},~{\boldsymbol h}_{k_j}[i]={\boldsymbol g}_{eq_{k_j}}[i]f_{k_j}[i]$ \\
 \qquad \qquad $\nabla P_{E_{{\boldsymbol w}_j}}[i]  = \frac{1}{M\sqrt {2\pi}\sqrt{{\boldsymbol w}^\emph{H}_j[i]{\boldsymbol w}_j[i]}}\sum^{M}_{j=1}\exp\big(-\frac{c^2_j[i]}{2}\big)sgn(s_j)
    \frac{{\boldsymbol{\bar{r}_j}}[i] - \frac{1}{2}\bar{s}_j[i]{\boldsymbol w}_j[i]}{\sigma_{n}{\boldsymbol w}^\emph{H}_j[i]{\boldsymbol w}_j[i]},$ \\
 \qquad \qquad $\nabla P_{E_{a_{S_j}}}[i] = \frac{1}{M\sqrt{2\pi}\sigma_{n}\sqrt{{\boldsymbol w}^\emph{H}_j[i]{\boldsymbol w}_j[i]}}\sum^{M}_{j=1}exp\big(-\frac{c^2_j[i]}{2}\big)sgn(s_j)\Re[{\boldsymbol w}^\emph{H}_j[i]{\boldsymbol h}_{SDA_j}[i]s_j],$ \\
 \qquad \qquad $\nabla P_{E_{a_{k_j}}}[i] = \frac{1}{M\sqrt{2\pi}\sigma_{n}\sqrt{{\boldsymbol w}^\emph{H}_{D_j}[i]{\boldsymbol w}_{D_j}[i]}}\sum^{M}_{j=1}exp\big(-\frac{c^2_j[i]}{2}\big)sgn(s_j)\Re[{\boldsymbol w}^\emph{H}_{D_j}[i]{\boldsymbol h}_{k_j}[i]s_j],$ \\
 \hline

 \qquad 2-3: \bfseries{JAPA SG MSR Algorithm} \\
 \qquad \qquad ${\boldsymbol R}_{SDA}[i] = {\boldsymbol H}^\emph{H}_{SDA}[i]{\boldsymbol W}[i]{\boldsymbol W}^\emph{H}[i]{\boldsymbol H}_{SDA}[i]{\boldsymbol A}_S[i],$ \\
 \qquad \qquad ${\boldsymbol R}_{{\boldsymbol G}_{eq_k}}[i] = {\boldsymbol G}^\emph{H}_{eq_{k}}[i]{\boldsymbol W}[i]{\boldsymbol W}^\emph{H}[i]{\boldsymbol H}_{SDA}[i]{\boldsymbol A}_S[i]{\boldsymbol A}_S^\emph{H}[i]{\boldsymbol F}_{SR_k}^\emph{H}[i],$ \\
 \qquad \qquad $n_{eq}[i] = \sigma^2_nTr({\boldsymbol W}^\emph{H}[i]{\boldsymbol W}[i]+{\boldsymbol W}^\emph{H}[i](\sum_{k=1}^{n_r}{\boldsymbol G}_{eq_k}[i]{\boldsymbol A}_k[i])(\sum_{k=1}^{n_r}{\boldsymbol G}_{eq_k}[i]{\boldsymbol A}_k[i])^\emph{H}{\boldsymbol W}[i]),$ \\
 \qquad \qquad $\nabla_{{\boldsymbol W}[i]} = \frac{\sigma^2_s}{n_{eq}[i]}(Tr(\parallel{\boldsymbol H}_{SDA}[i]{\boldsymbol A}_S[i]\parallel^2_F{\boldsymbol W}[i])n_{eq}[i]$ \\
  \qquad \qquad \qquad \qquad $-\parallel{\boldsymbol W}^\emph{H}[i]{\boldsymbol H}_{SDA}[i]{\boldsymbol A}_S[i]\parallel^2_F Tr({\boldsymbol W}[i]+{\boldsymbol G}_{eq_k}[i]{\boldsymbol A}_k[i]({\boldsymbol G}_{eq_k}[i]{\boldsymbol A}_k[i]){\boldsymbol W}[i])),$ \\
 \qquad \qquad $\nabla_{a_{S_j}[i]} = \frac{\sigma^2_s}{n_{eq}[i]}r_{SDA_j}[i],$ \\
 \qquad \qquad $\nabla_{a_{R_kD}[i]} = \frac{\sigma^2_s}{n_{eq}[i]}(r_{{\boldsymbol G}_{eq}}[i]n_{eq}[i] - \sigma^2_nr_{{{\boldsymbol N}_k}_j}[i]\sum_{j=1}^{N}r_{{\boldsymbol G}_{eq_{k_j}}}[i]a^*_{k_j}[i]),$ \\
 ~~~ \bfseries {end for}\\
  \hline
3: \bfseries{Update}: \\
 \qquad \qquad ${\boldsymbol w}_j[i + 1] = {\boldsymbol w}_j[i] - \mu\nabla_{{\boldsymbol w}^*_j[i]},$ \\
 \qquad \qquad $a_{S_j}[i + 1] = a_{S_j}[i] - \nu\nabla_{a^*_{S_j}[i]},$ \\
 \qquad \qquad $a_{k_j}[i + 1] = a_{k_j}[i] - \tau\nabla_{a^*_{k_j}[i]},$ \\
 \hline
4: \bfseries{Normalization}: \\
 \qquad \qquad ${\boldsymbol A}_S[i+1] = \frac{\sqrt{\rm{P_T}}{\boldsymbol A}_S[i+1]}{\parallel{\boldsymbol A}_S[i+1]\parallel_F},$ \\
 \qquad \qquad ${\boldsymbol A}_k[i+1] = \frac{\sqrt{\rm{P_R}}{\boldsymbol A}_k[i+1]}{\parallel\sum_{k=1}^{n_r}\rm{Tr}({\boldsymbol A}_k[i+1]\parallel_F}.$ \\

\hline

    \end{tabular}
    \end{small}
\end{table}

\section{Analysis}

The proposed JAPA SG algorithms employ three different criteria to
compute the power allocation matrices iteratively at the destination
node and then send them back via a feedback channel. In this
section, we will illustrate the low computational complexity
required by the proposed JAPA SG algorithms compared to the existing
power allocation optimization algorithms using the same criteria and
will examine their feedback requirement.

\subsection{Computational Complexity Analysis}

In Table II, we compute the number of additions and multiplications
to compare the complexity of the proposed JAPA SG algorithms with
the conventional power allocation strategies. The computational
complexity of the proposed algorithms is calculated by summing the
number of additions and multiplications, which is related to the
number of antennas $N$, the number of relay nodes $n_r$, and the $N
\times T$ STC scheme employed in the network. Note that the
computational complexity in \cite{Ref1} and \cite{Ref2} is high
because the key parameters in the algorithms can only be obtained by
eigenvalue decomposition, which requires a high-cost computing
process when the matrices are large \cite{Keke}.

\begin{table}
  \centering
  \caption{Computational Complexity of the Algorithms}\label{}
  \begin{tabular}{ccc}
  \toprule
  & \multicolumn{2}{c}{Number of operations per symbol}\\
  \cmidrule{2-3}
  Algorithm & Multiplications & Additions \\
  \midrule
  PA MMSE(III-A) & $(T+1)^6N^6+(T+1)N+8(T+1)N$ & $7(T+1)N+2$ \\
  JAPA MMSE SG(IV-A) & $(7T+5)N$ & $4(T+1)N$ \\
  JAPA MBER SG(IV-B) & $(M+1)(T+1)N+M$ & $(2M+1)(T+1)N$ \\
  OPA\cite{Ref2} & $N^4+2N^2+N^2T^2$ & $2NT-1$\\
  JAPA MSR SG(IV-C) & $7(T+1)N+N+1$ & $7(T+1)N+N+2$ \\
  PO-PR-SIM\cite{Ref1} & $N^4+2N^2$ & $2NT$\\
  \bottomrule
\end{tabular}
\end{table}

\subsection{Feedback Requirements}

The proposed JAPA SG algorithms require communication between the
relay nodes and the destination node according to different
algorithms. The feedback channel we considered is modeled as an AWGN
channel. A $4$-bit quantization scheme, which quantizes the real
part and the imaginary part by $4$ bits, respectively, is utilized
prior to the feedback channel. More efficient schemes employing
vector quantization \cite{gersho,delamare_ieeproc} and that take
into account correlations between the coefficients are also
possible.

For simplicity we show how the feedback errors in power allocation
matrices at the relay nodes affect the accuracy of the detection and
only one relay node is employed. The $N \times N$ diagonal power
allocation matrix with feedback errors at the $k$th relay node is
derived as
\begin{equation}\label{6.1}
    {\boldsymbol {\hat{A}}}[i] = {\boldsymbol A}[i] + {\boldsymbol E}[i],
\end{equation}
where ${\boldsymbol A}[i]$ denotes the accurate power allocation matrix and ${\boldsymbol E}[i]$ stands for the error matrix. We assume the parameters in ${\boldsymbol E}[i]$ are Gaussian with zero mean and variance $\sigma_f$. Then the received symbol vector is given by
\begin{equation}\label{6.2}
\begin{aligned}
    {\boldsymbol {\hat{r}}}[i] & = {\boldsymbol G}_{eq}[i]{\boldsymbol {\hat{A}}}[i]{\boldsymbol F}[i]{\boldsymbol s}[i]+{\boldsymbol G}_{eq}[i]{\boldsymbol {\hat{A}}}[i]{\boldsymbol n}_{SR}[i]+{\boldsymbol n}_{RD}[i]\\
    & = {\boldsymbol G}_{eq}[i]{\boldsymbol {\hat{A}}}[i]{\boldsymbol F}[i]{\boldsymbol s}[i]+{\boldsymbol {\hat{n}}}_D[i],
\end{aligned}
\end{equation}
where ${\boldsymbol {\hat{n}}}_D[i]$ denotes the received noise with zero mean and variance $\sigma_f({\boldsymbol I}+\parallel{\boldsymbol G}_{eq}[i]{\boldsymbol {\hat{A}}}[i]\parallel_F)$. By defining ${\boldsymbol {\hat{p}}} = E \left[{\boldsymbol {\hat{r}}}{\boldsymbol s}^\emph{H}\right]$ and ${\boldsymbol {\hat{Rx}}}=E \left[{\boldsymbol {\hat{r}}}{\boldsymbol {\hat{r}}}^\emph{H}\right]$, we can obtain the MSE with the feedback errors as
\begin{equation}\label{6.4}
\begin{aligned}
    m_e & = \rm{Tr}({\boldsymbol {\hat{p}}}^\emph{H}{\boldsymbol {\hat{Rx}}}^{-1}{\boldsymbol {\hat{p}}})\\
    & = \rm{Tr}\big(({\boldsymbol G}_{eq}[i]({\boldsymbol A}[i] + {\boldsymbol E}[i]){\boldsymbol F}[i]\sigma_s)^\emph{H}(\parallel{\boldsymbol G}_{eq}[i]({\boldsymbol A}[i] + {\boldsymbol E}[i]){\boldsymbol F}[i]\parallel^2_F\sigma_s+({\boldsymbol I}+\parallel{\boldsymbol G}_{eq}[i]({\boldsymbol A}[i] + {\boldsymbol E}[i])\parallel^2_F)\sigma_f)^{-1}\\
    &~~~~~~~~({\boldsymbol G}_{eq}[i]({\boldsymbol A}[i] + {\boldsymbol E}[i]){\boldsymbol F}[i]\sigma_s)\big),
\end{aligned}
\end{equation}
while the MSE expression of the system with accurate power allocation parameters is given by
\begin{equation}\label{6.5}
    m = \rm{Tr}\big(({\boldsymbol G}_{eq}[i]{\boldsymbol A}[i]{\boldsymbol F}[i]\sigma_s)^\emph{H}(\parallel{\boldsymbol G}_{eq}[i]{\boldsymbol A}[i]{\boldsymbol F}[i]\parallel^2_F\sigma_s+({\boldsymbol I}+\parallel{\boldsymbol G}_{eq}[i]{\boldsymbol A}[i]\parallel^2_F)\sigma_n)^{-1}({\boldsymbol G}_{eq}[i]{\boldsymbol A}[i]{\boldsymbol F}[i]\sigma_s)\big).
\end{equation}
By substituting (\ref{6.5}) into (\ref{6.4}), we can obtain the difference between the MSE expressions with accurate and inaccurate power allocation matrices which is given by
\begin{equation}\label{6.6}
\begin{aligned}
    m_e & = m + \rm{Tr}\big(({\boldsymbol G}_{eq}[i]{\boldsymbol E}[i]{\boldsymbol F}[i]\sigma_s)^\emph{H}(\parallel{\boldsymbol G}_{eq}[i]{\boldsymbol E}[i]{\boldsymbol F}[i]\parallel^2_F\sigma_s+({\boldsymbol I}+\parallel{\boldsymbol G}_{eq}[i]{\boldsymbol E}[i]\parallel^2_F)\sigma_n)^{-1}({\boldsymbol G}_{eq}[i]{\boldsymbol E}[i]{\boldsymbol F}[i]\sigma_s)\big)\\
    & = m + m_{eo}.
\end{aligned}
\end{equation}
The received power allocation matrices are positive definite
according to the power constraint, which indicates $m_{eo}$ is a
positive scalar. The expression in (\ref{6.6}) denotes an analytical
derivation of the MSE at the destination node, which indicates the
impact of the limited feedback employed in the JAPA SG algorithms.

\section{Simulations}

The simulation results are provided in this section to assess the
proposed JAPA SG algorithms. The equal power allocation (EPA)
algorithm in \cite{Yindi Jing} is employed in order to identify the
benefits achieved by the proposed power allocation algorithms. The
cooperative MIMO system considered employs an AF protocol with the
Alamouti STBC scheme in \cite{Tong1} using BPSK modulation in a
quasi-static block fading channel with AWGN. The effect of the
direct link is also considered. It is possible to employ the DF
protocol or use a different number of antennas and relay nodes with
a simple modification. The ML detection is considered at the
destination node to indicate the achievement of full receive
diversity even though other detection algorithms
\cite{choi,windpassinger,delamaretvt} can also be adopted. The
system is equipped with $n_r=1$ relay node and $N=2$ antennas at
each node. In the simulations, we set the symbol power $\sigma^2_s$
to 1. The $SNR$ in the simulations is the received $SNR$ which is
calculated by (\ref{4.9}).

\begin{figure}
\begin{center}
\def\epsfsize#1#2{0.825\columnwidth}
\epsfbox{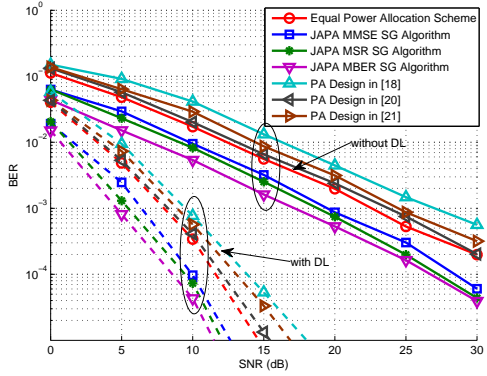} \caption{SNR versus BER for JAPA SG
Algorithms}\label{p2} \vspace{-1em}
\end{center}
\end{figure}

The proposed JAPA SG algorithms derived in Section IV are compared
with the EPA algorithm and the power allocation algorithms in
\cite{Ref1}, \cite{WeiG} and \cite{Munoz} with and without the
direct link (DL) in Fig. 2. The results illustrate that the
performance of the proposed JAPA SG algorithms is superior to the
EPA algorithm by more than $3$dB. The performance of the power
allocation algorithms in the literature are designed for AF systems
without re-encoding at the relays and in order to obtain a fair
comparison, they have been adapted to the system considered in Fig.
1. However, as shown in the plot, the performance of the existing
power allocation algorithms cannot achieve a BER performance as good
as the proposed algorithms. In the low SNR scenario, the JAPA MSR SG
algorithm can achieve a better BER performance compared with the
JAPA MMSE SG algorithm, while with the increase of the SNR, the BER
curves of the JAPA MSR and MMSE SG algorithms approach the BER
performance of the JAPA MBER SG algorithm with enough Monte-Carlo
simulation numbers. The BER of the JAPA MBER SG algorithm achieves
the best performance because of the received BER is minimized by the
algorithm in Section IV. The performance improvement of the proposed
JAPA SG algorithms is achieved with more relays employed in the
system as an increased spatial diversity is provided by the relays.

\begin{figure}
\begin{center}
\def\epsfsize#1#2{0.825\columnwidth}
\epsfbox{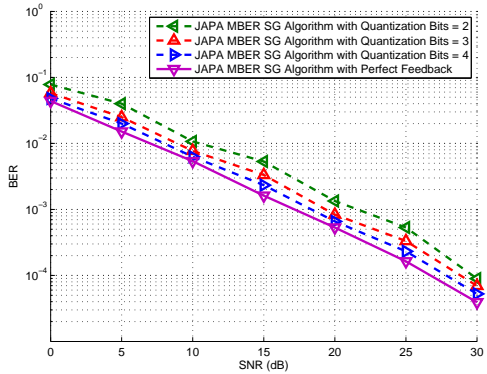} \caption{JAPA MBER SG Algorithm SNR versus
BER}\label{p3} \vspace{-1em}
\end{center}
\end{figure}

The simulation results shown in Fig. 3 illustrate the influence of
the feedback channel on the JAPA MBER SG algorithm. As mentioned in
Section V, the optimized power allocation matrices will be sent back
to each relay node and the source node through an AWGN feedback
channel. The quantization and feedback errors are not considered in
the simulation results in Fig. 2, so the optimized power allocation
matrices are perfectly known at the relay node and the source node
after the JAPA SG algorithm; while in Fig. 3, it indicates that the
performance of the proposed algorithm will be affected by the
accuracy of the feedback information. In the simulation, we use
$2,~3,~4$ bits to quantize the real part and the imaginary part of
the element in ${\boldsymbol A}_S[i]$ and ${\boldsymbol A}_k[i]$,
and the feedback channel is modeled as an AWGN channel. As we can
see from Fig. 3, by increasing the number of quantization bits for
the feedback, the BER performance approaches the performance with
perfect feedback, and by making use of $4$ quantization bits for the
real and imaginary part of each parameter in the matrices, the
performance of the JAPA SG algorithm is about 1dB worse.

\begin{figure}
\begin{center}
\def\epsfsize#1#2{0.825\columnwidth}
\epsfbox{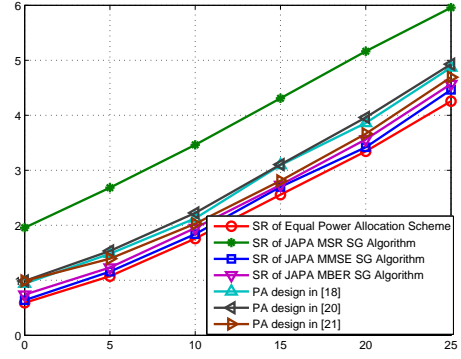} \caption{JAPA SG Algorithms Sum Rate versus
SNR}\label{p4} \vspace{-1em}
\end{center}
\end{figure}

The transmission rate of the cooperative MIMO network with EPA and
PA schemes in \cite{Ref1}, \cite{WeiG} and \cite{Munoz} and the
proposed JAPA SG algorithms in Section IV-C is given by Fig. 4. The
number of relay nodes is equal to $1$ for all the algorithms. The
proposed JAPA MSR SG optimization algorithm adjusts the power
allocated to each antenna in order to achieve the maximum of the sum
rate in the system. From the simulation results, it is obvious that
a higher throughput can be achieved by the existing PA algorithms in
\cite{Ref1}, \cite{WeiG} and \cite{Munoz} compared to the proposed
JAPA MMSE and MBER SG algorithms. The reason for that lies in the
design criterion of the existing and the proposed algorithms.
However, the improvement in the sum rate by employing the JAPA MSR
SG algorithm can be observed as well. The rate improvement of the
JAPA MMSE and MBER SG algorithms is not as much as the JAPA MSR SG
algorithm because the optimization of the proposed JAPA MMSE and
MBER optimization algorithms are not suitable for the maximization
of the sum rate.

\begin{figure}
\begin{center}
\def\epsfsize#1#2{0.825\columnwidth}
\epsfbox{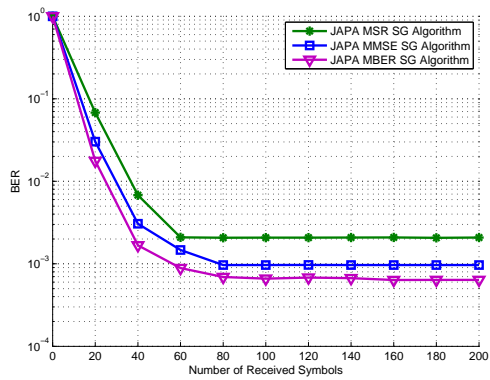} \caption{BER performance vs. Number of Symbols
for JAPA SG Algorithms}\label{p5} \vspace{-1em}
\end{center}
\end{figure}

The simulation results shown in Fig. 5 illustrate the convergence
property of the proposed JAPA SG algorithm. All the schemes have an
error probability of $0.5$ at the beginning, and after the first
$20$ symbols are received and detected, the JAPA MMSE scheme
achieves a better BER performance compared with the JAPA MSR scheme
and the JAPA MBER scheme a better BER than the other algorithms.
With the number of received symbols increasing, the BER curve of all
the schemes are almost straight, while the BER performance of the
JAPA MBER algorithm can be further improved and obtain a fast
convergence after receiving $80$ symbols.

\section{Conclusion}

We have proposed joint adaptive power allocation and receiver design
algorithms according to different criteria with the power constraint
between the source node and the relay nodes, and between relay nodes
and the destination node to achieve low BER performance. Joint
iterative estimation algorithms with low computational complexity
for computing the power allocation parameters and the linear receive
filter have been derived. The simulation results illustrated the
advantage of the proposed power allocation algorithms by comparing
it with the equal power allocation algorithm. The proposed algorithm
can be utilized with different DSTC schemes and a variety of
detectors \cite{vikalo} \cite{delamare_spadf} and estimation
algorithms \cite{jidf} in cooperative MIMO systems with AF strategy
and can also be extended to the DF cooperation protocols.

\appendices



\ifCLASSOPTIONcaptionsoff
  \newpage
\fi

\bibliographystyle{IEEEtran}

\end{document}